\documentclass[aps, prd, tightenlines, notitlepage, superscriptaddress, twocolumn, nofootinbib, preprintnumbers, floatfix, showkeys,10pt]{revtex4-2}
\usepackage{graphicx}
\usepackage{orcidlink}
\usepackage[normalem]{ulem}
\usepackage{amssymb}
\usepackage{amsmath}
\usepackage{graphicx}
\usepackage{url}
\usepackage{ulem}
\usepackage[utf8]{inputenc}
\usepackage{slashed}
\usepackage{lipsum}
\usepackage[T1]{fontenc}
\usepackage{appendix}
\usepackage{csquotes}
\usepackage[x11names]{xcolor}
\usepackage{cleveref}

\pdfoutput = 1

\RequirePackage[normalem]{ulem}

\crefname{section}{Sec.}{Secs.}
\crefname{figure}{Fig.}{Figs.}
\crefname{equation}{Eq.}{Eqs.}
\crefname{table}{Table}{Tables}
\crefname{appendix}{Appendix}{Appendices}

\begin{document}

\bibliographystyle{utphys}

\title{Spin Identification of Dark Sector Mediators through Angular Distributions}

\author{D. Aristizabal Sierra\orcidlink{0000-0001-5429-3708}}%
\email{daristizabal@uliege.be}%
\affiliation{Universidad T\'ecnica Federico Santa Mar\'{i}a-Departamento de F\'{i}sica\\Casilla 110-V, Avda. Espa\~na 1680, Valpara\'{i}so, Chile}%

\author{S. Fuenzalida Garrido\orcidlink{0000-0002-7835-5157}}%
\email{sebastian.fuenzalidg@usm.cl}%
\affiliation{Universidad T\'ecnica Federico Santa Mar\'{i}a-Departamento de F\'{i}sica\\Casilla 110-V, Avda. Espa\~na 1680, Valpara\'{i}so, Chile}%

\author{F. Kling\,\orcidlink{0000-0002-3100-6144}}%
\email{fkling@uni-bonn.de}%
\affiliation{Universit\"at Bonn, Regina-Pacis-Weg 3, D-53113 Bonn, Germany}%

\author{T. M\"akel\"a\orcidlink{0000-0002-1723-4028}}%
\email{tmakela@uci.edu}%
\affiliation{Department of Physics and Astronomy, University of California, 92697, Irvine, CA, USA}%

\author{N. Viaux\orcidlink{}}
\email{nicolas.viaux@usm.cl}
\affiliation{Universidad T\'ecnica Federico Santa Mar\'{i}a-Departamento de F\'{i}sica\\Casilla 110-V, Avda. Espa\~na 1680, Valpara\'{i}so, Chile}

\begin{abstract}
A variety of experiments are operating or planned to search for displaced decays of light long-lived dark sector particles. In case such a state is discovered, the next step is determining its quantum numbers. We identify an angular observable, reconstructible solely from the decay products' four-momenta, that exhibits an anisotropic distribution for vector bosons from light meson decays and an isotropic distribution for scalars. We demonstrate that searches at DUNE, SHiP and FASER2 will be able to identify the mediator spin in sizable regions of yet unconstrained parameter space.
\end{abstract}

\maketitle


\textbf{Introduction.} Dark sectors with MeV-scale dark matter (DM) are well-motivated theoretical scenarios that address the origin of the DM~\cite{Pospelov:2007mp, Jaeckel:2010ni}. In their simplest realizations the Standard Model (SM) gauge group is extended with an extra $U(1)_D$ symmetry, that can be either global or local. In realizations with a global $U(1)_D$, the spectrum involves a pseudo-Nambu-Goldstone boson if the symmetry is spontaneously broken (see e.g.~\cite{Peccei:1977hh, Weinberg:1977ma, Wilczek:1977pj, Wilczek:1982rv}). Axion-like particles are well-known examples of these new degrees of freedom. A different category of models contain local $U(1)_D$ symmetries. In these cases, the spectrum contains a new gauge boson, which kinetically mixes with the photon and thereby obtains a feeble coupling to SM fermions~\cite{Okun:1982xi, Holdom:1985ag, delAguila:1988jz}. The dark sector may comprise dark scalars as well, related with the spontaneous symmetry breaking of dark sector symmetries. In all these types of models, interactions between the dark and visible sector are mediated by these degrees of freedom. This \enquote{bridging} fixes the DM relic density, providing the means to test these hypotheses at the laboratory level.

Dark sector mediators, along with DM, can be abundantly produced in collider experiments. In most searches a high-energy proton beam is impinged on a high-density fixed target~\cite{Batell:2009di}. Dark sector particles can be produced either directly or indirectly, from the hadron activity triggered by the collision itself. SHiP and NA62 rely on this strategy~\cite{Alekhin:2015byh, NA62:2017rwk}. Neutrino accelerator facilities, although not motivated by dark sector searches, operate with the same strategy. Experiments of this type involve proton beam energies spanning two orders of magnitude. COHERENT and experiments at Fermilab, in particular the upcoming DUNE experiment, are rather good examples~\cite{COHERENT:2021pvd, COHERENT:2022pli, DUNE:2020fgq, DUNE:2021tad}. The same production strategy applies at the LHC, but in $pp$ collisions. FASER and FASER2 are designed following this approach~\cite{Feng:2017uoz, FASER:2018eoc, FASER:2022hcn, Salin:2927003, FPF:2025bor, Adhikary:2024nlv, FPFWorkingGroups:2025rsc}.

Search strategies depend heavily on mediator decay topologies. Particularly relevant for current and forthcoming long-lived particle experiments are visible decay modes: the new state would be produced in an unmonitored high intensity collision point, travels a macroscopic distance, and decays after in a dedicated displaced vertex detector. The particular decay topology depends on the available phase space and model realization. \enquote{Clean} signals are displaced lepton-pair resonances, searched for with tracking systems. SHiP, NA62, FASER(2), and DUNE, along with other dedicated proposals, have excellent projected sensitivities \cite{Alekhin:2015byh, NA62:2017rwk, Feng:2017uoz, DUNE:2020fgq, DUNE:2021tad, CODEX-b:2019jve, HIKE:2022qra, AristizabalSierra:2026jgp}. 

The identification of a dark sector mediator signal, with fully visible clean topologies, follows from the reconstruction of the lepton-pair invariant mass. That information alone allows the determination of the mediator mass. Furthermore, measurements of the decay length, if possible and available, will enable the extraction of the new interaction coupling. However, a full deconstruction of the underlying physics model at work demands as well the determination of the mediator spin.

In this letter, we focus on dark photons produced in pseudoscalar mesons decays $\pi^0, \eta, \eta^{\prime} \to \gamma A'$ decaying to fermion pairs $A' \to e^+ e^-$ and demonstrate that angular distribution measurements in the mediator reference frame allow the determination of the mediator spin. In this regard, meson decays provide an observable rather suited for such a task. In the dark photon-mediated process, since the initial state carries no angular momentum only the dark photon transverse polarization degrees of freedom contribute. Thus, the electron (or positron) angular distribution is anisotropic in the mediator frame. On the contrary, if the observed particle is a scalar or pseudoscalar, the angular distribution is instead expected to be isotropic. 

For the anisotropy to be apparent the dark photon polarization degrees of freedom should be properly taken into account, i.e. spin correlations should be included~\cite{Feng:2025gji}. This alone, however, does not suffice. For the anisotropy to be reconstructable it should be expressed in terms of a measurable angle. We identify such an angle and determine the 95\% CL regions where the electron anisotropy can be measured. We specialize our results for existing (NA62) and forthcoming (FASER2, DUNE and SHiP) facilities and comment on requirements needed to measure such anisotropy. \medskip


\textbf{Theoretical framework.} The uniting feature of light dark sector models is that they contain a light mediator that couples both to the SM sector and dark sector. Two concrete examples are a scalar mediator $\phi$ (the dark Higgs, related with dark sector spontaneous symmetry breaking) \cite{Pospelov:2008zw,Patt:2006fw} and a vector mediator $A'$ (the dark photon) \cite{Okun:1982xi,Holdom:1985ag,delAguila:1988jz}, whose interactions to SM fermions $f$ are
\begin{equation}
    \label{Lag}
    \mathcal{L}_\phi = \theta \frac{m_f}{v} \bar f f \phi
    \quad\text{and}\quad
    \mathcal{L}_{A'} = \epsilon q_f \bar f \gamma^\mu f A'_\mu\ .
\end{equation}
Here $\theta$ determines visible-dark scalar mixing, $v\simeq 246\,$GeV, $\epsilon$ is the kinetic mixing parameter, and $q_f$ is the fermion charge. For sufficiently heavy DM masses, the mediator will exclusively decay to SM particles. If the relevant couplings $\theta$ and $\epsilon$ are small, these mediators are long-lived and decay at a macroscopic distance away from their production.

\renewcommand{\arraystretch}{1.5}
\setlength{\tabcolsep}{3.5mm}
\begin{table*}[t]
    \centering
    \begin{tabular}{c c|c c c c}
        \hline\hline
        Facility
        & Experiment
        & Energy
        & Event Statistics
        & $d$ $[\text{m}]$
        & $L$ $[\text{m}]$\\
        \hline
        LHC
        & FASER2
        & $\sqrt{s}=13.6$ TeV
        & $\mathcal{L}=3000$ fb$^{-1}$
        & $620$
        & $10$ \\
        LBNF
        & DUNE
        & $E_p = 120$~GeV
        & $N_\text{POT}=1.47\times 10^{22}$
        & $574$
        & $10$ \\
        SPS
        & NA62
        & $E_p = 400$~GeV
        & $N_\text{POT}=10^{18}$  
        & $105$
        & $75$ \\
        SPS
        & SHiP
        & $E_p = 400$~GeV
        & $N_\text{POT}=6\times 10^{20}$  
        & $60$
        & $50$\\
        \hline\hline
    \end{tabular}
    \caption{Collider and fixed target experimental geometry parameters. Distance $d$ refers to values measured from the interaction point to the most upstream part of the fiducial decay volume. Detector length $L$ to the length of the fiducial decay volume.}
    \label{tab:detector_details}
\end{table*}
In case of an observation of such mediator decays in one of the dedicated long-lived-particle experiments, a natural question is whether the underlying physics model can be deconstructed. The mediator mass and its coupling to SM fermions can be inferred with the aid of invariant mass and decay lengths measurements. The mediator spin, however, requires a different strategy.

In Ref.~\cite{Feng:2025gji}, Feng. et al have studied dark photon production in pion decay $\pi \to \gamma A'$, followed by the dark photon decay $A' \to e^+ e^-$. They identified an angular observable $\theta^*$, defined as the angle between the electron momentum in the dark photon rest frame and the dark photon momentum in the pion rest frame, that is sensitive to the mediator spin. For a scalar mediator, the probability distribution is isotropic, $\rho(\cos\theta^*) = 1/2$, while for a vector mediator the distribution is anisotropic
\begin{align}
    \rho (\cos\theta^*) = \frac{3}{8} \frac{(1 \!+\! 4r^2) + (1\!-\!4r^2) \cos^2\theta^*}{1 \!+\! 2 r^2}
    \ .
    \label{eq:correlation1}
\end{align}
Here $r=m_f/m$, where $m$ is the dark photon mass and $m_f$ is the final-state fermion mass. The distribution has only a slight dark photon mass dependence, relevant in regions where $m\simeq m_f$. For larger values of the dark photon mass one finds $\rho\simeq 3/8 \cdot(1+\cos^2\theta^*)$.  Note that reconstructing this observable requires knowledge on the full event kinematics. This, however, is in general not possible in long-lived particle searches, where only the mediator decay products are observed.

\begin{figure}[h!]
    \centering
    \includegraphics[scale=0.6]{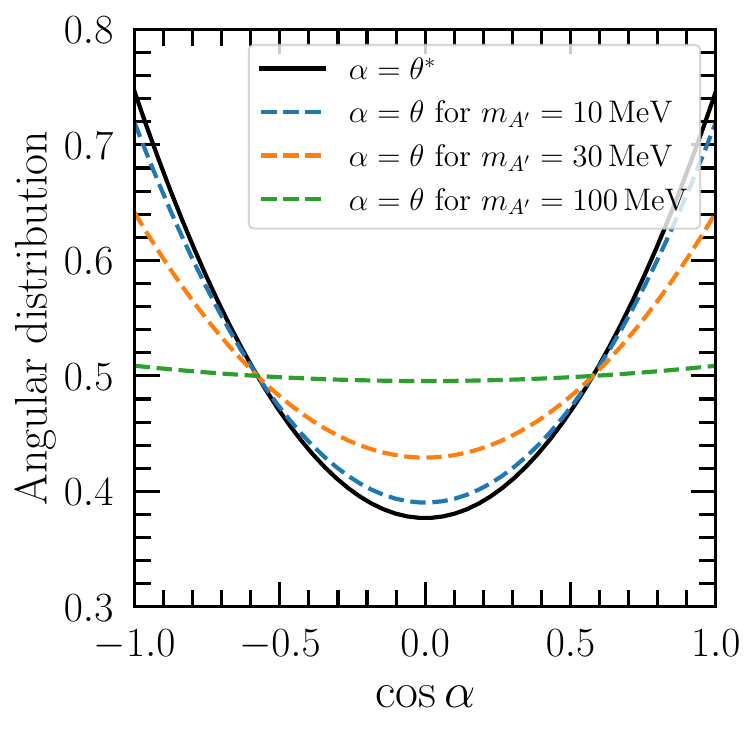}
    \caption{Electron angular probability distribution in terms of $\theta$ and $\theta^*$ for $\pi^0 \to \gamma A'$ and $A' \to e^+ e^-$.}
    \label{fig:lab_dist}
\end{figure}
Ignoring spin correlations will render the dark photon probability distribution isotropic as well. The fact that only the dark photon transverse polarization degrees of freedom contribute, means that the anisotropy is of dynamical origin rather than a kinematic effect. This implies that a certain level of anisotropy should be present regardless of the \enquote{quantization} axis employed for the measurement of the electron polar angle in the dark photon frame.

Thus, one can calculate the electron angular probability distribution in terms of an angle that can be reconstructed in terms of laboratory-frame kinematic variables. The angle between the electron momentum in the dark photon frame and the dark photon momentum in the laboratory frame is such variable
\begin{equation}
    \label{eq:costheta}
    \cos\theta=\frac{2\varepsilon_q}{m }
    \frac{\vec{p}_\cdot\widehat{q}-|\vec{q}|/\varepsilon_q\,E_-}{\sqrt{|\vec{q}|^2 - 4m_e^2}}\ ,
\end{equation}
where $\varepsilon_q=E_-+E_+$ and $\vec{q}=\vec{p}_-+\vec{p}_+$, with $E_\pm$ and $\vec{p}_\pm$ the electron and positron total energy and momentum. In terms of this kinematic variable we find an isotropic angular distribution for a scalar mediator, while for a dark photon
\begin{align}
    \rho(\cos\theta) = \frac{3}{8} \frac{[1\!+\!\mathcal{F}] \!+\! 4r^2[1\!-\!\mathcal{F}] \!+\! [1\!-\!4r^2][1\!-\!3\mathcal{F}] \cos^2\!\theta}{1 + 2 r^2},
    \label{eq:correlation2}
\end{align}
where
\begin{align}
    \label{eq:F_fun}
    \!\!\mathcal{F} \!=\! \frac{4 M^2 m^2} {(M^2\!-\!m^2)^2} \!\left( \!\frac{\text{atanh}[(M^2\!\!-\!m^2)/(M^2\!\!+\!m^2)]}{(M^2\!-\!m^2)/(M^2\!+\!m^2)} \!-\!1 \! \right)\ .
\end{align}
Here $M$ refers to the initial-state meson mass. It should be emphasized that in contrast to the distribution in Eq. \eqref{eq:correlation1}, this expression depends strongly on the dark photon mass in the all available parameter space. Thus, the reconstruction of the distribution not only provides information on the mediator spin but also on its mass. A few representative examples of the distribution are shown in \cref{fig:lab_dist}, along with the result obtained in terms of $\theta^*$. The behavior can be readily understood.  When $m/M\to 0$ (but with $m>2m_e$), $\mathcal{F}\to 0$ and the distribution matches the result in terms of $\theta^*$. On the contrary, when $m/M\to 1$, $\mathcal{F}\to 1/3$ and the angular dependence goes away.

Technical details of the derivation of Eq. \eqref{eq:correlation2} are presented in App.~\ref{app:derivation}. Qualitatively, however, the result can be easily understood: The momentum vector of the laboratory is isotropically distributed in the pion rest frame. After boosting into the dark photon frame, it therefore follows a dipole distribution that is peaked in the direction of the pion momentum. This dipole is stronger for large boosts (small dark photon masses). Therefore, for small dark photon masses, the laboratory momentum points in the same direction as the pion momentum, and hence $\theta \approx \theta^*$. The behavior of Eq. \eqref{eq:correlation2} at $\mathcal{F}\to 0$ is a manifestation of this effect. In contrast, for heavy dark photons, the laboratory momentum vector is almost isotropically distributed in the dark photon rest frame, and therefore $\cos\theta$ is randomly distributed. Note that this result is independent of the dark photon energy in the laboratory frame. This means that one can recover the same distribution function also for a dark photon beam with a broad energy distribution.

\begin{figure*}[t]
    \centering
    \includegraphics[scale=0.6]{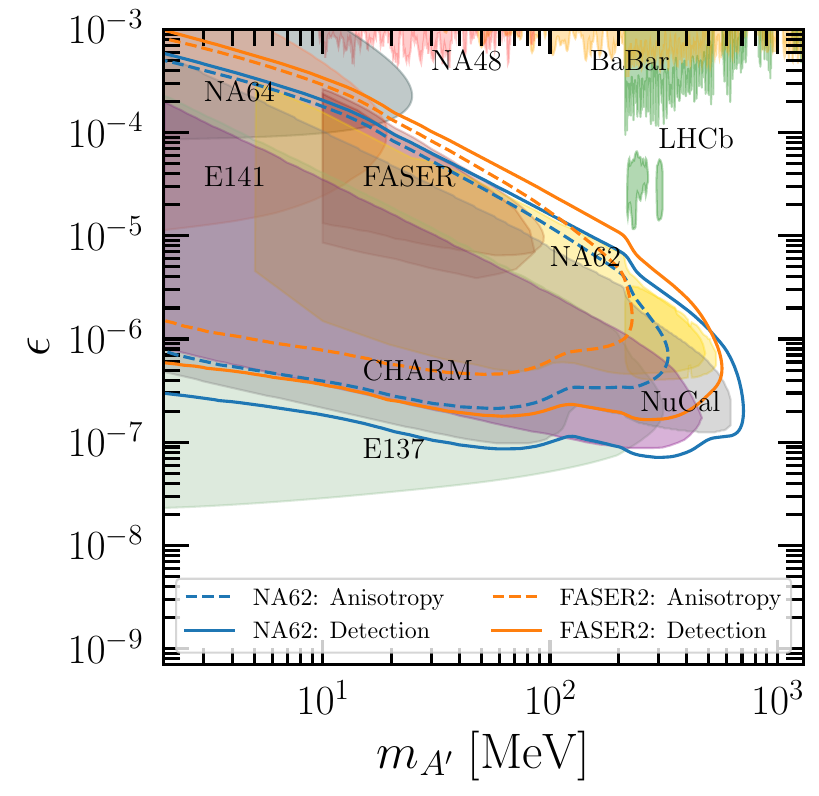}
    \includegraphics[scale=0.6]{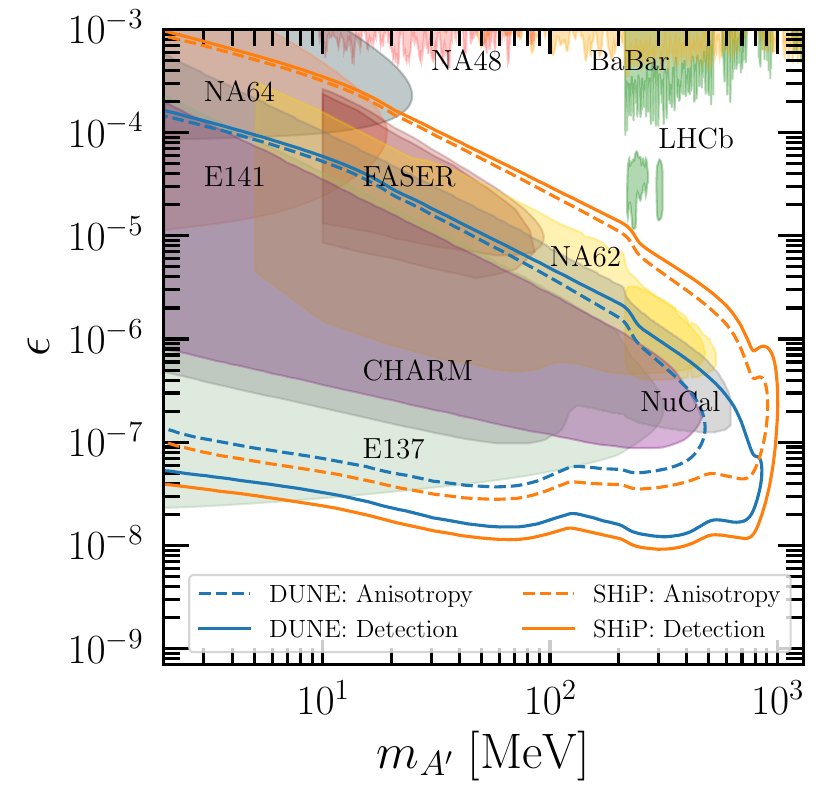}
    \caption{95\% CL anisotropy sensitivity contours from $e^-$ angular distribution measurements, along with current most relevant exclusion limits. Shown as well are discovery reach 95\% CL contours (detection in the plot legends).}
    \label{fig:dark_photon_reach}
\end{figure*}

Finally, let us note that \cref{eq:correlation2} allows spin correlations to be included in dark photon signal simulators in a simple way. In such simulators, the production and decay are typically simulated separately. The results derived here show how spin correlations, despite the fact that they result from the interplay between production and decay, can be incorporated solely by modifying the way in which dark photons are decayed, without modifying how they are produced. \medskip


\textbf{Projected Sensitivities.} Having identified suitable observables, we now investigate the sensitivity of several experiments to identify the spin of a long-lived particle. The determination of anisotropy sensitivities relies on a log-likelihood analysis. The null hypothesis is defined by the angular distribution expectation from a scalar mediator (isotropic emission). The alternative hypothesis is instead defined by the angular anisotropy expected from a dark photon mediator. Since the angular probability distributions are in both cases known, $1/2$ for the scalar case and Eq.~\eqref{eq:correlation2} for the dark photon case, how much they differ can be quantified analytically. Details can be found in App.~\ref{app:stat_sig}.

Assuming a single-event likelihood ratio and running over a $10^3\times 10^3$ grid in the $m_{A^\prime}$ versus $\epsilon$ parameter space, we derive 95\% CL expected sensitivity intervals for NA62, FASER2, DUNE and SHiP. Using the light meson spectra obtained from \texttt{EPOSLHC}~\cite{Pierog:2013ria}, we estimate the event rate of dark photons decaying inside the detectors fiducial volumes with \texttt{FORESEE}~\cite{Kling:2021fwx}. Detector lengths as well as baselines and collision variables are displayed in Tab.~\ref{tab:detector_details}. Further details can be found in Refs. \cite{Berryman:2019dme, DUNE:2020fgq, ship, faser2, NA62:2023nhs}. In all cases we assume that measurements are background free, not affected by systematic uncertainties, and that there is no other exotic production mechanism at work. Thus, our results can be regarded as best-case sensitivities.

Results of the analysis are shown in Fig. \ref{fig:dark_photon_reach}. For comparison, results for the discovery reach are displayed too along with the most relevant current exclusion limits. For the region of interest they include NA64~\cite{NA64:2016oww, NA64:2018lsq, NA64:2023wbi, Gninenko:2018ter}, NuCal~\cite{Blumlein:1990ay, Blumlein:2011mv}, NA62~\cite{NA62:2023nhs}, CHARM~\cite{CHARM:1985an, Blumlein:2011mv}, E137~\cite{Bjorken:1988as, Batell:2014mga}, and FASER~\cite{FASER:2023tle, FASER:2024bbl}. Note that the discovery reach is based solely on event counting, while the measurement of the anisotropy requires a comparison between two hypotheses that predict rather different outputs. Thus, the latter requires an event discrimination that the former does not (up to possible background). The discovery covers larger regions in parameter space because of this reason.

We find that NA62 does not have a sufficient event rate to determine the dark photon spin. In contrast, FASER2 and DUNE will be able to perform such a measurement in yet unexplored regions of parameter space at $\epsilon \sim 10^{-4}$ and $10^{-7}$, respectively. Due to its large collision rate and volume, SHIP was found to have the best expected sensitivity extending over larger portions of unconstrained parameter space. \medskip

\begin{figure}[h]
\includegraphics[scale=0.6]{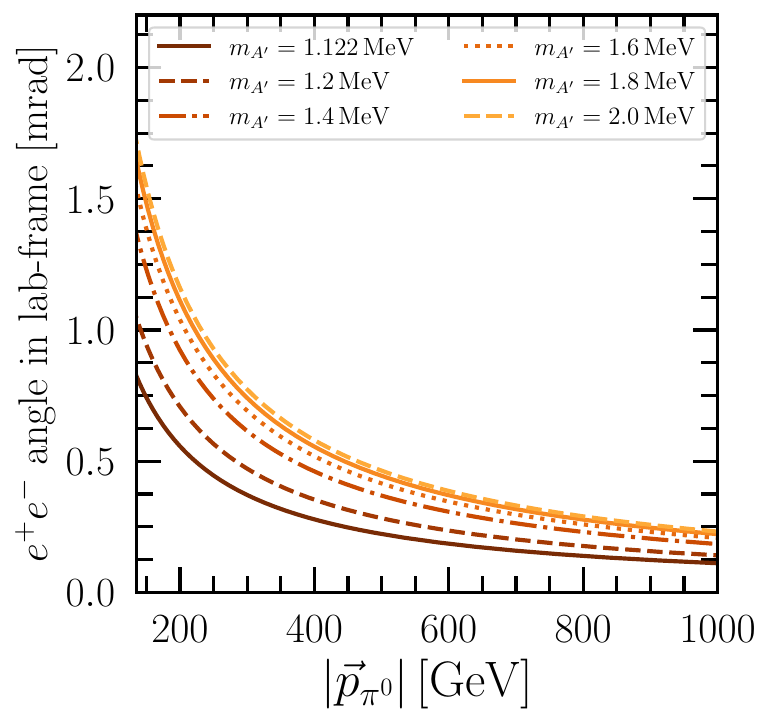}
\caption{The lab-frame angle between the $e^+e^-$ pair as a function of the initial $\pi^0$ momenta for select dark photon masses. For a 0.6~TeV $\pi^0$ beam, this implies a required angular resolution of 0.2~mrad for $>2$~MeV dark photon masses.}
\label{fig:lepton_angle_estimate}
\end{figure}

\textbf{Detector angular resolution requirements.} Experimental resolution affects the measured individual lepton momenta, and the angle $\theta_\pm$ between them in the laboratory frame. As a first estimate of how much angular smearing renders the polarized and unpolarized cases indistinct, consider a meson moving along the $z$-axis in the laboratory frame and radiating a collinear dark photon with mass $m_{A'}$, subsequently decaying to $e^+ e^-$. In the dark photon rest frame, the $e^+e^-$ pair is produced back-to-back in the $xy$-plane. The leptons receive transverse momenta of $\pm\vec{p}_\perp$, and energies of $m_{A'}/2$, and are boosted to the laboratory frame by $\beta_z$ corresponding to the initial meson $z$ momentum. Their opening angle becomes
\begin{equation}
\label{eq:opening_angle}
\theta_\pm = \arccos\left(\frac{1-r'}{1+r'}\right)\ ,
\end{equation}
with $r'$ given by
\begin{equation}
\label{eq:r_parameter}
r' \equiv 
\left(\frac{|\vec{p}_\perp|}{\gamma\beta_z E_e}\right)^2 =
\frac{1 - 4 m_e^2/m_{A'}^2}{\gamma^2 \beta_z^2}\ .
\end{equation}
If the angular smearing width is greater than $\theta_\pm/2$, the vector directions become essentially randomized. As the boost to the laboratory frame is defined by the meson momentum and not the dark photon, $\gamma$ and $\beta$ are independent of $m_{A'}$. Thus the required angular resolution has only a slight dependence on $m_{A'}$, when close to $m_e$, and the result becomes $m_{A'}$ independent at $m_{A'}>2$~MeV, as illustrated in Fig.~\ref{fig:lepton_angle_estimate} for $\pi^0$. Hence the experimental resolution requirements are expected to be approximately constant for most of the parameter space of interest here.

The impact of experimental angular and energy resolution is assessed in further detail by simulating the production of $e^+e^-$ pairs in the decays of a dark photon with and without spin correlations, corresponding to meson momenta characteristic of selected experiments. The event samples are produced using \texttt{FORESEE} \cite{Kling:2021fwx}, and Gaussian smearing with incremental widths is applied on the energies and directions of the leptons in each generated event to determine if the $\theta_\pm$ distributions for the two cases may become indistinguishable. The smearing of energies is found to have no effect on shapes of the $\theta_\pm$ distributions, it only affects the accuracy of the reconstruction of the dark photon mass. The variable of interest is therefore the maximal amount of angular smearing with which the two spectra remain distinct. 

For characteristic initial pion energies of 10-100~GeV (1~TeV), the cases with and without spin correlations are found to be well distinguishable with an angular resolution of $\sim$0.1~mrad ($\sim$50~$\mu$rad). This is close to the optimistic estimate of \cref{fig:lepton_angle_estimate}. Note that, although reconstructing the highest energy events is challenging, simultaneously measuring both lepton energies with high accuracy is not necessary here. Since $\delta \theta \approx \delta x / L$, with $\delta x$ the tracker position resolution and $L$ the tracker length, the results imply that e.g. a $\sim 10$~m long detector should be designed to resolve $\delta x \sim 0.1$~mm, setting the spatial resolution a detector must reach to measure the anisotropy.


\textbf{Conclusions.} A variety of experiments are operating or planned to search for displaced decays of light long-lived particles, such as dark photons or dark scalars. In case of a discovery of such a new state, the natural next step will be to determine its spin. Motivated by this question, in this letter we studied these decays including spin correlations. We have demonstrated that the electron angular distribution in the dark photon frame is anisotropic, compared with the scalar case that follows an isotropic distribution. Therefore, measurements of the angular distribution in that frame provide a mean for the identification of the mediator spin.

A meaningful statement, however, requires expressing the angular probability distribution in terms of measurable quantities. We have accomplished this by identifying an angle that can be entirely reconstructed in terms of the lepton-pair momenta measured in the laboratory frame: the angle between the electron momentum in the dark photon frame and the dark photon momentum in the laboratory frame. With this result we have studied the sensitivity of existing and forthcoming experiments: NA62, FASER2, DUNE and SHiP. We have shown that the latter will be able to measure angular anisotropies in sizable regions of parameter space. Finally, we commented on the experimental requirements needed to reliably reconstruct this observable. \medskip

\textbf{Acknowledgments.} The work of D.A.S. is supported by ANID grant \enquote{Fondecyt Regular} 1260595. The work of T.M. is supported in part by U.S.~National Science Foundation Grants PHY-2111427 and PHY-2210283 and Heising-Simons Foundation Grant 2020-1840. The work of S.F.G is supported by ANID grant \enquote{Fondecyt de Postdoctorado} 3260317. The work of N.V. is supported by the Millennium Institute for Subatomic Physics at the High Energy Frontier (SAPHIR), ANID Millennium Science Initiative Program ICN2019\_044.

\bibliography{biblio}

\clearpage
\onecolumngrid

\appendix

\begin{figure*}[b!]
    \centering
    \includegraphics[scale=0.9]{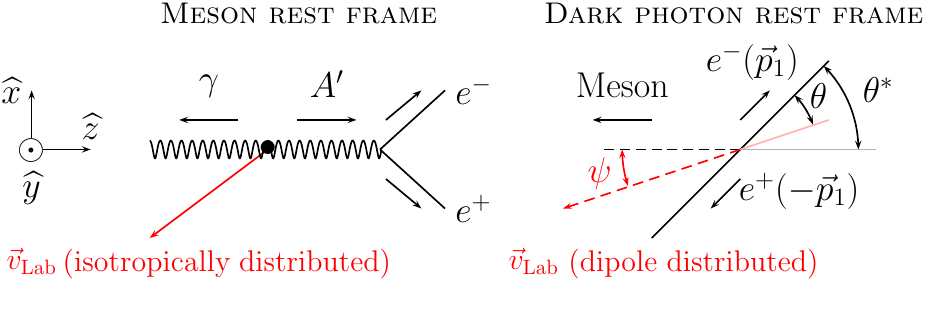}
    \caption{Relevant vectors in the meson and dark photon rest frames used in the calculation of the electron angular distribution in terms of $\theta$.}
    \label{fig:kinematics}
\end{figure*}

\section{Derivation of the electron misalignment angle}
\label{app:derivation}

In this appendix, we calculate the probability disribution for the angular observable $\cos\theta$, as given by Eqs. \eqref{eq:correlation2} and \eqref{eq:F_fun}. To start with we first consider the kinematics of the problem as illustrated in Fig.~\ref{fig:kinematics}. Shown in the left hand side is the decay in the meson ($\pi^0,\eta^0,\eta^\prime$) rest frame: The meson (with mass $M$) decays into a dark photon (with mass $m$), which we choose to be emitted along the $z$-direction. From the kinematics of the problem one finds
\begin{equation}
    \label{eq:meson_frame_kinematics}
    E = \frac{M^2+m^2}{2M}\ ,\qquad
    |\vec{p}|=\frac{M^2-m^2}{2M}\ .
\end{equation}
The dark photon velocity and boost factor are expressed in terms of these kinematic variables according to $\beta=|\vec{p}|/E$ and $\gamma=E/m$. In that frame, the dark photon angular distribution is isotropic. Thus, $\vec{v}_\text{Lab}$, refering to the laboratory velocity in that frame, and $\vec{p}$, the dark photon momentum, are not correlated. This means that $\vec{v}_\text{Lab}$ is isotropically distributed.

Let us now perform a boost along the $z$ direction and move into the dark photon rest frame, which is illustrated in the right side of Fig.~\ref{fig:kinematics}. The meson momentum now points in the negaive $z$ direction and the momentum of the electron (with mass $m_f$) points in the direction $\hat p_1 = (0,\sin\theta^*, \cos\theta^*)$. Here we have introduced the angle $\theta^*$ between the electron in the dark photon rest frame (so $\vec{p}_1$) and the dark photon momentum in the pion restframe (so the z-axis). This is the angle for which the paper by Feng et al~\cite{Feng:2025gji} calculated the distribution function presented in Eq.~\eqref{eq:correlation1}.

Reconstructing the electron angular distribution in terms of $\theta^*$ requires measuring the parent meson momentum, which is not accesible in long-lived particle experiments. An alternative angular variable $\theta$, shown in the same graph, is defined as the angle between the electron momentum in the dark photon rest frame and the dark photon momentum in the laboratory frame (equivalently, the negative laboratory momentum in the dark photon rest frame). 

To proceed with the calculation of the electron distribution in terms of $\theta$, we write the laboratory velocity in the dark photon rest frame according to $-\widehat{v}_\text{Lab}=(\sin\psi\cos\phi,\sin\psi\sin\phi,\cos\psi)$, with $\psi$ and $\phi$ the polar and azimuthal angles, respectively. A key observation is that while the azimuthal distribution is still isotropic, $\psi$ follows a dipole distribution instead, as a consequence of a relativistic beaming effect induced by the boost along $z$. Thus, we write
\begin{equation}
    \label{eq:dipole_dist}
    \rho(\cos\psi, \phi) = \rho(\cos\psi) = \frac{1}{4 \pi \gamma^2 (1+\beta \cos\psi)^2}\ .
\end{equation}
Quantitatively, the angle $\theta$ can be understood as the misalignment between $\widehat{p}_1$ and $\widehat{v}_1$. This means that
\begin{equation}
    \label{eq:misalignment}
    \cos\theta = -\widehat{v}_\text{Lab} \cdot \hat{p}_1 = \sin\theta^* \sin\psi \sin\phi + \cos\theta^* \cos\psi
\end{equation}

Note that the kinematic limits of the angular distribution $\rho(\cos\theta)$ can be derived solely using the kinematic criteria, Eqs.~\eqref{eq:meson_frame_kinematics}. In the extreme non-relativisitic limit, $m=M$, the dark photon and laboratory frames match, implying that $\vec{v}_\text{Lab}$ is isotropically distributed. Since there is no preferred direction anymore, $\rho(\cos\theta)=1/2$. The other extreme case is the ultra-relativistic limit $m\ll M$. In that case, in the dark photon rest frame, $-\vec{v}_\text{Lab}=z$ and thus $\theta=\theta^*$ (see right graph in Fig. \ref{eq:meson_frame_kinematics}). For the distribution one then expects $\rho(\cos\theta)=\rho(\cos\theta^*)$. These limiting cases will allow the validation of the full distribution.

From Eq. \eqref{eq:misalignment} one can see that
$\rho(\cos\theta)$ should depend on $\theta^*$, $\psi$ and $\phi$. The new angular probability distribution is therefore a joint probability distribution subject to the constrain dictated by Eq. \eqref{eq:misalignment}. It can thus be written according to
\begin{equation}
     \rho(\cos\theta) = \int \rho(\cos\theta^*)  \rho(\cos\psi, \phi)\delta(\cos\theta - \sin\theta^* \sin\psi \sin\phi - \cos\theta^* \cos\psi) d\cos\theta^* d\cos\psi d\phi\ .
\end{equation}
Noting that $\rho(\cos\psi, \phi)$ is independent of $\phi$, we can easily perform the integral over $\phi$ by using 
\begin{equation}
    \label{eq:useful_rel}
    \int \delta(A - B \sin\phi) d\phi = 2\frac{\Theta(B^2-A^2)}{\sqrt{B^2-A^2}}\ ,
\end{equation}
with $A,B$ general constants with respect to $\phi$, and the $\Theta$-function ensuring the existence of a solution. We obtain
\begin{equation}
    \rho(\cos\theta) = \int \rho(\cos\theta^*)  \rho(\cos\psi) \frac{2 \Theta[(\sin\theta^* \sin\psi)^2 - (\cos\theta-\cos\theta^* \cos\psi)^2]}{\sqrt{(\sin\theta^* \sin\psi)^2 - (\cos\theta-\cos\theta^* \cos\psi)^2} } \  d\cos\theta^* \ d\cos\psi\ ,
\end{equation}
which can be rewritten in terms of the variables $u = \cos\theta^*$, $v=\cos\theta$, and $t=\cos\psi$ as follows
\begin{equation}
    \label{eq:dist_before_int}
    \rho(v) = 2 \int \rho(u)  \rho(t) \frac{\Theta[(1-u^2)(1-v^2) - (t-uv)^2]}{\sqrt{(1-u^2)(1-v^2) - (t-uv)^2} }  du \ dt\ .
\end{equation}
Eq. \eqref{eq:dist_before_int} is the general solution to the problem. Integration, however, is not straghtforward. Thus, before analyzing the general case we first consider the kinematic limit $m=M$.
\subsection{Special Case $M=m$} 
\label{sec:special_case_m_eq_M}
Let us, for a moment, consider the special case in which $m=M$ (so $\beta=0$ and $\gamma=1$). In that limit, the distribution in \eqref{eq:dipole_dist} reduces to $1/4\pi$. In this case we find
\begin{equation}
    \rho(v) 
    = \int \frac{\rho(u) du}{2\pi} \int^{1}_{-1} \frac{ \Theta[(1-u^2)(1-v^2) - (t-uv)^2]}{\sqrt{(1-u^2)(1-v^2) - (t-uv)^2} } dt 
    =\int \frac{\rho(u) du}{2\pi} \int^{s(t=1)}_{s(t=-1)} \frac{ \Theta[1-s^2]}{\sqrt{1-s^2} } ds\ . 
\end{equation}
Here we introduced $s = (t-uv)/\sqrt{(1-u^2)(1-v^2)}$. One can easily check that $s(t=-1)<-1$ and $s(t=1)>1$. However, the $\Theta$-function forces $|s|<1$. This allows us to simplify 
\begin{equation}
    \int^{s(t=1)}_{s(t=-1)} \frac{ \Theta[1-s^2]}{\sqrt{1-s^2} } ds 
    =\int^{1}_{-1} \frac{ 1}{\sqrt{1-s^2} } ds 
    =\Big[ \text{arcsin(s)}\Big]^{1}_{-1}
    = \pi\ .
\end{equation}
We thus finally obtain 
\begin{equation}
    \rho{\cos\theta} = \rho(v) 
    =\int \frac{\rho(u) du}{2\pi} \int^{s(t=1)}_{s(t=-1)} \frac{ \Theta[1-s^2]}{\sqrt{1-s^2} } ds 
    = \frac{\pi}{2\pi}\int \frac{\rho(u) du}{2\pi} 
    = \frac{1}{2}\ .
\end{equation}
This implies that there is no anisotropy in the distribution of $\theta$, independent of whether there is one in $\theta^*$.

\subsection{General Case}
\label{sec:general_case}
Let us go back to the general case dictated by Eq. \eqref{eq:dipole_dist} without further assumptions. We can write
\begin{equation}
    \rho(v) 
    = \int \frac{\rho(u) du}{2 \pi \gamma^2}  \int^{1}_{-1} \frac{1}{(1+\beta t)^2}\frac{ \Theta[(1-u^2)(1-v^2) - (t-uv)^2]}{\sqrt{(1-u^2)(1-v^2) - (t-uv)^2} } dt\ .
\end{equation}
In terms of the variable $s$ used in the Sec. \ref{sec:special_case_m_eq_M} and the argument on the integration boundaries, we obtain 
\begin{equation}
    \rho(v) 
    = \int \frac{\rho(u) du}{2 \pi \gamma^2}  \int^{1}_{-1} \frac{1}{(A +  B s )^2} \frac{ 1}{\sqrt{1-s^2} } ds\ ,
\end{equation}
where we have defined
\begin{equation}
    \label{eq:A_and_B}
    A = 1+\beta u v
    \qquad\text{and}\qquad
    B = \beta \sqrt{(1-u^2)(1-v^2)}\ .
\end{equation}
We can solve the integral over $s$ analytically and obtain 
\begin{equation}
\begin{aligned}
    \label{eq:rho_v_general_case}
     \int^{1}_{-1} \frac{1}{(A+Bs)^2}\frac{ ds}{\sqrt{1-s^2}} 
     = \frac{- 2 A  }{(A^2 - B^2)^{3/2} } 
     \left[\tan^{-1}\left( \frac{s \sqrt{A^2 - B^2}}{ A + B s}\right)\right]_{-1}^{1}\ .
\end{aligned}
\end{equation}
Using the relation $\tan^{-1}(x) - \tan^{-1}(y) = \tan^{-1}((x-y)/(1+xy))$, the term in brackets turns out to be
\begin{equation}
     \left[\tan^{-1}\left( \frac{s \sqrt{A^2 - B^2}}{ A + B s}\right)\right]_{-1}^{1}  
     = \tan^{-1}\left( \frac{\sqrt{A^2 - B^2}}{ A + B}\right) -  \tan^{-1}\left( \frac{\sqrt{A^2 - B^2}}{ B - A}\right) 
     = 
     \tan^{-1}(\infty) =  - \frac{\pi}{2}\ ,
\end{equation}
where we used the fact that $1+xy = 1+ \frac{\sqrt{A^2 - B^2}}{ A + B} \frac{\sqrt{A^2 - B^2}}{ B-A} = 1-\frac{A^2-B^2}{B^2-A^2}=0$. Inserting this result back in Eq. \eqref{eq:rho_v_general_case}, we can write
\begin{equation}
     \int^{1}_{-1} \frac{1}{(A+Bs)^2}   \frac{ ds}{\sqrt{1-s^2}} 
     =  \frac{\pi A}{(A^2-B^2)^{3/2}} 
     =  \frac{\pi (1+\beta u v)}{((1+\beta u v)^2-\beta^2 (1-u^2)(1-v^2))^{3/2}}\ ,
\end{equation}
and thus 
\begin{equation}
    \rho(v) 
    = \int \frac{\rho(u) du}{2 \pi \gamma^2}  \int^{1}_{-1} \frac{1}{(A +  B s )^2} \frac{ 1}{\sqrt{1-s^2} } ds  
    = \int \frac{\rho(u) du}{2  \gamma^2} 
    \frac{ (1+\beta u v)}{((1+\beta u v)^2-\beta^2 (1-u^2)(1-v^2))^{3/2}}\ .
\end{equation} 

\noindent \textbf{Isotropic Case:} Let us first consider the case of an isotropic distribution $\rho(u)=1/2$. In this case, we obtain for the distribution 
\begin{equation}
\begin{aligned}
    \rho(v) 
    &= \int \frac{du}{4  \gamma^2} 
    \frac{ (1+\beta u v)}{((1+\beta u v)^2-\beta^2 (1-u^2)(1-v^2))^{3/2}} 
    = \frac{1}{4  \gamma^2} \left[ \frac{ \beta v + u}{(\beta^2-1)\sqrt{\beta^2 (u^2+v^2-1) + 2 \beta u v + 1} }\right]_{-1}^{1} \\
    &= \frac{1}{4  \gamma^2} \left[ \frac{ \gamma^2(\beta v + u)}{\sqrt{\beta^2 v^2 + 2 \beta u v + 1} }\right]_{-1}^{1}  
    = \frac{1}{4 } \left( \frac{ \beta v + 1}{\sqrt{\beta^2 v^2 + 2 \beta  v + 1} } -  \frac{ \beta v - 1}{\sqrt{\beta^2 v^2 - 2 \beta  v + 1} } \right) 
    = \frac{1}{2}\ .
\end{aligned}
\end{equation}
As expected, this provides an isotropic distribution. \medskip

\begin{figure}[t]
     \centering
     \includegraphics[scale=0.5]{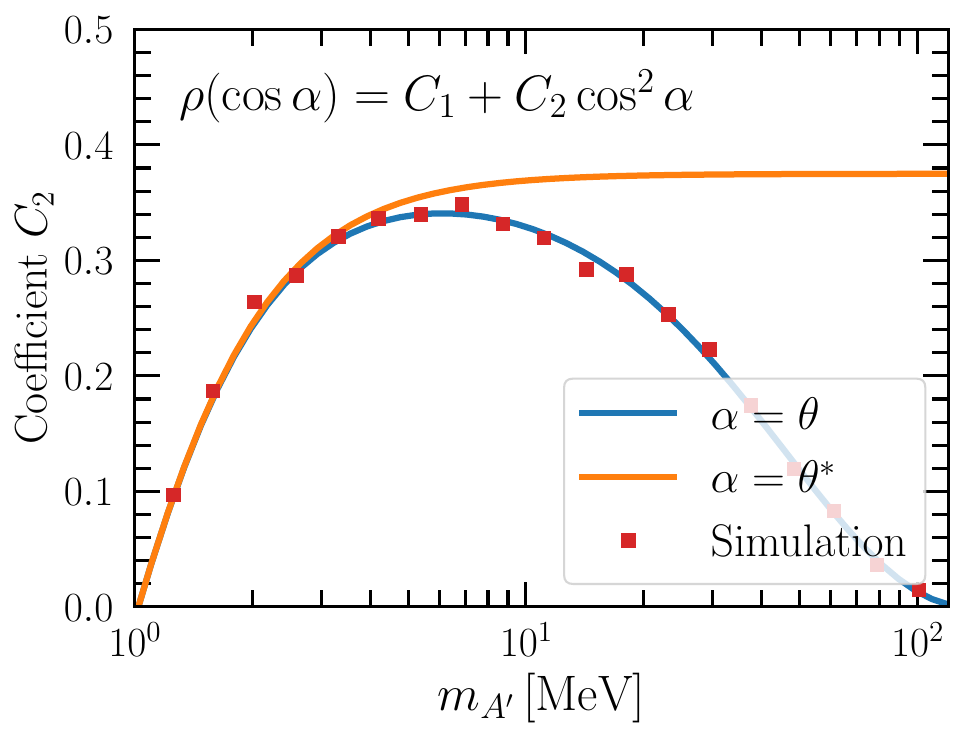}
     \caption{Coefficient $C_2$ as a function of a few dark photon mass values obtained from a full simulation (red squares). Results are inline with our analytical findings for $\rho(\cos\theta)$, Eq. \eqref{eq:final}.}
    \label{fig:angular}
\end{figure}

\noindent \textbf{Anisotropic Case:} Let us now consider the case of an anisotropic distribution. Using the short-hand notation $C_1=(m^2 + 4m_f^2)/(m^2 + 2 m_f^2)$, and $C_2=(m^2-4m_f^2)/(m^2 + 2 m_f^2)$, we can write Eq.~\eqref{eq:correlation1} as $\rho(u) = C_1 + C_2 u^2$. We have already calculated
\begin{equation}
    \label{eq:I1}
   I_1 = \int_{-1}^{1} \frac{du}{2  \gamma^2} 
     \frac{ (1+\beta u v)}{((1+\beta u v)^2-\beta^2 (1-u^2)(1-v^2))^{3/2}}  = 1\ .
\end{equation}
Thus, let us now calculate the integral 
\begin{equation}
\begin{aligned}
    \label{eq:I2}
    I_2 (\beta, v)&= \int_{-1}^{1} \frac{ du}{2  \gamma^2} 
    \frac{ u^2 (1+\beta u v)}{((1+\beta u v)^2-\beta^2 (1-u^2)(1-v^2))^{3/2}}  \\
    &= \frac{1}{2\gamma^2} \frac{2\beta( (1-\beta^2)(1-3v^2) - \beta^2v^2) + (1-\beta^2)(3v^2-1)   \left[\text{asinh}\left(\frac{\beta - v}{\sqrt{(1-\beta^2)(1-v^2)}}\right) +\text{asinh}\left(\frac{\beta + v}{\sqrt{(1-\beta^2)(1-v^2)}}\right)
    \right]}{\beta^3(\beta^2 - 1)}\\
    & = \frac{1}{2\gamma^2} \frac{2\beta( (1-\beta^2)(1-3v^2) - \beta^2v^2) + 2 (1-\beta^2)(3v^2-1) \cdot \text{atanh}(\beta)}{\beta^3 (\beta^2 - 1)} \\
    & = -\frac{ (1-3v^2) - \beta^2v^2 \gamma^2 +  (3v^2-1) \cdot \text{atanh}(\beta)/\beta}{\beta^2\gamma^2} 
    = \frac{1}{\beta^2\gamma^2} \left[\left( \frac{\operatorname{atanh}(\beta)}{\beta} -1 \right) + v^2\left(3 + \beta^2\gamma^2 - 3\frac{\operatorname{atanh}(\beta)}{\beta}\right) \right]\ .\\
\end{aligned}
\end{equation}
Combining the results in Eqs. \eqref{eq:I1} and \eqref{eq:I2}, we get the distribution function
\begin{equation}
\begin{aligned}
    \rho(v)  
    &= \int \frac{(C_1 + C_2 u^2) \ du}{2  \gamma^2} 
    \frac{ (1+\beta u v)}{((1+\beta u v)^2-\beta^2 (1-u^2)(1-v^2))^{3/2}} 
    =  C_1 I_1 + C_2 I_2 (\beta, v) \\
    &= \left[ C_1 + \frac{C_2} {\beta^2 \gamma^2}\left( \frac{\text{atanh}(\beta)}{\beta} -1 \right) \right] + \left[\frac{C_2} {\beta^2 \gamma^2} \left(3 + \beta^2\gamma^2 - 3\frac{\operatorname{atanh}(\beta)} {\beta}\right) \right]v^2\\
    &= \left[ C_1 + \frac{C_2} {\beta^2 \gamma^2}\left( \frac{\text{atanh}(\beta)}{\beta} -1 \right) \right] + \left[C_2 \left(1 - \frac{3}{\beta^2\gamma^2} \left( \frac{\operatorname{atanh}(\beta)} {\beta} -1\right) \right) \right]v^2\ .
\end{aligned}
\end{equation}
Inserting back our origial notation, we get our final answer 
\begin{equation}
\begin{aligned}
    &\rho(\cos\theta) \!=\! \frac{3}{8} \frac{m^2+4m_f^2}{m^2 + 2 m_f^2}\left[1 \!+\! \frac{m^2-4m_f^2}{m^2+4m_f^2}\frac{4 M^2 m^2} {(M^2\!-\!m^2)^2}\mathcal{F} \right] + \frac{3}{8} \frac{m^2-4m_f^2}{m^2 + 2 m_f^2} \left[1\!-\!\frac{12 M^2 m^2} {(M^2\!-\!m^2)^2} \mathcal{F} \right] \cos^2\!\theta \\
    &\text{with}\quad \mathcal{F} \!=\! \frac{\text{atanh}[(M^2\!-\!m^2)/(M^2\!+\!m^2)]}{(M^2\!-\!m^2)/(M^2\!+\!m^2)} -1 \ .
    \label{eq:final}
\end{aligned}
\end{equation}
One can easily check that $\int_{-1}^1 \rho(v) dv = 1$. This prediction agrees well with the simulation, as shown in Fig.~\ref{fig:angular}. Here we obtained the angular spectra for a set of dark photon masses, fitted them with a template $\rho(\cos\theta) = C_1 + C_2 \cos^2\theta$, and plotted the coefficent $C_2$ versus the dark photon mass $m$. The prediction of Eq.~\eqref{eq:final} perfectly describes the simulated data. 

\section{Statistical Significance}
\label{app:stat_sig}
The observable of interest for our discussion is $x=\cos\theta$, where $-1<x<1$. In the null-hypothesis (no anisotropy) the single event likelihood function is flat $p_0(x) = 1/2$. In the alternative hypothesis (with anisotropy), the single event likelihood function is given by $p_1(x) = \left(\frac 1 2-\frac {C_2}{3} \right) + C_2 \cdot x^2$. Following the results of the previous section, we have 
\begin{equation}
    C_2(m) =  \frac{3}{8}\frac{m^2-4m_f^2}{m^2 + 2 m_f^2}
    \left[1\!-\!\frac{12 M^2 m^2} {(M^2\!-\!m^2)^2}  \left(\frac{\text{atanh}[(M^2\!-\!m^2)/(M^2\!+\!m^2)]}{(M^2\!-\!m^2)/(M^2\!+\!m^2)} -1\right) 
    \right]\ .
    \label{eq_appendix_statistical_sig:c_parameter}
\end{equation}
The full likelihood function $p_i^\text{full}(\{x\})$ for a set of events $\{x\}$, consiting of $n$ events can be written in terms of a Poisson term $\text{Pois}(n,N_i)$ for the observed number of events $n$ and expected number of events $N_i$, and a product of single event likelihoods $p(x_j)$ for each event $x_j$. We can write  
\begin{equation}
 p_i^\text{full}(\{x\}) = \text{Pois}(n,N_i) \times \prod_{j} p_i(x_j)\ .
\end{equation}
Using that both hypotheses lead to the same number of events, $N_0=N_1$, the full log-likelihood ratio $\log \left[r^\text{full}(\{x\})\right]$ between the two hypotheses can be written as a sum over the single event log-likelihoods 
\begin{equation}
\log \left[r^\text{full}(\{x\})\right] = \sum_j \log \frac{ p_0(x_j) }{ p_1(x_j)}\ .
\end{equation}

\begin{figure}[t]
    \centering
    \includegraphics[scale=0.47]{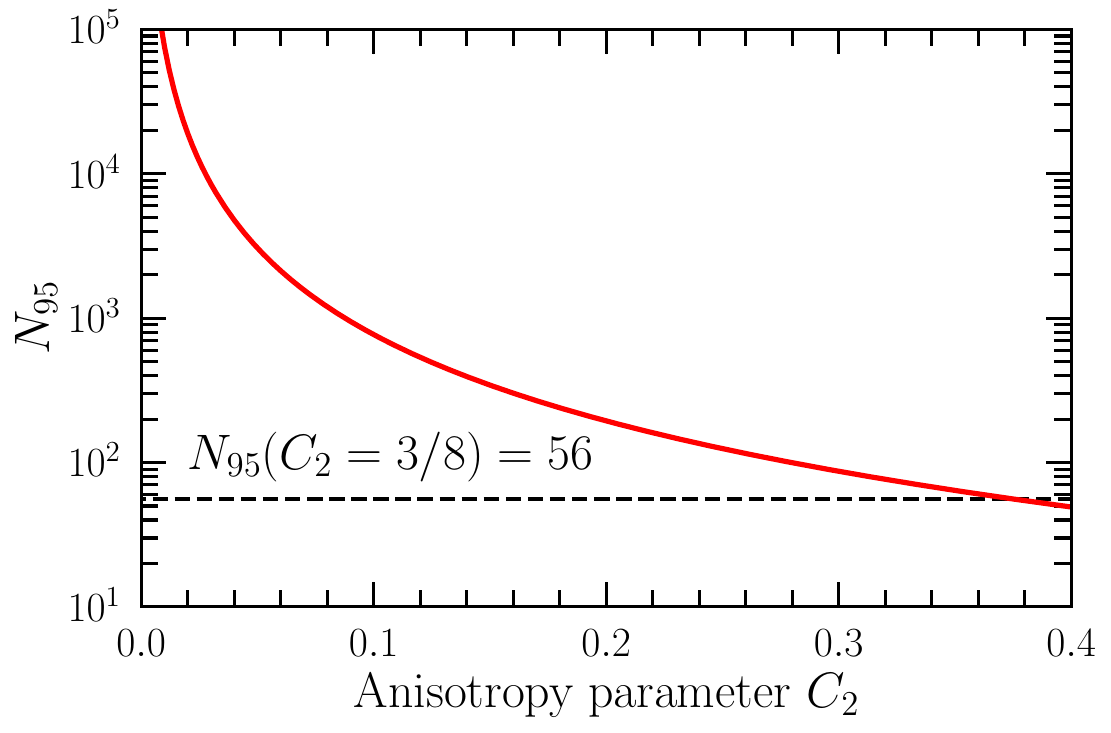}
    \includegraphics[scale=0.47]{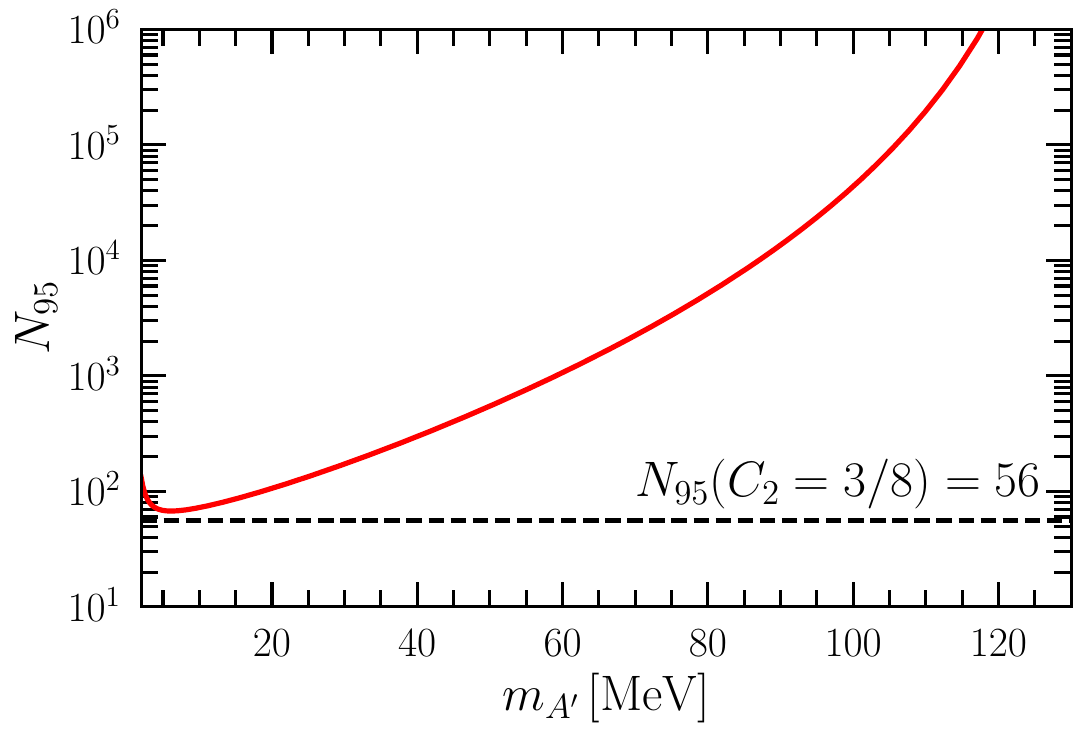}
    \caption{Number of events needed to reject the null hypotheses at 95\% CL, $N_{95}$, as a function of the anisotropy parameter $C_2$ and the dark photon mass $m$ for the decay chain $\pi^0 \to \gamma A'$ and $A' \to e^+ e^-$.}
    \label{fig:n95}
\end{figure}

Assuming the alternative hypothesis is correct, we would like to know how many events we need to exclude the null-hypothesis at a certain confidence level. We can calculate the expected log-likelilood ratio as
\begin{equation}
\!\!\mathcal{L} = E[\log\left[r^\text{full}(\{x\})\right]|p_{1}(x)] = N \int dx \log \left[\frac{ p_0(x) }{ p_1(x)}\right] p_1(x) = - N \left[\log 2 + \int dx\, \log p_1(x)\times p_1(x) \right] = - N I(C_2)\ ,
\end{equation}
which can be evaluated analytically, namely
\begin{equation}
\begin{aligned}
I(C_2)  &=  \log 2 + \int_{-1}^1 dx \int dx  
\log \left[\frac 1 2 - \frac {C_2} 3  + C_2\cdot x^2\right] \times [\frac 1 2 - \frac {C_2} 3 + C_2\cdot x^2] \\
&=  \log \left(1 + \frac {4C_2} {3} \right)
- \frac{4}{3}
+ \frac{4C_2}{9}
+ \frac{4(3 - 2C_2)}{9}
\sqrt{\frac{3 - 2C_2}{6C_2}}
\arctan \Big(
\sqrt{\frac{6C_2}{3 - 2C_2}}
\Big)\ .
\label{eq_appendix_statistical_sig:expected_number_of_events}
\end{aligned}
\end{equation}
Since we know that $-2 N(C_2) I(C_2) = -2 \mathcal{L}=Z^2$, we can estimate the number of events $N$ needed to reject the null hypothesis. To obtain a one-sided 95\% exclusion (common for excluding the null-hypothesis), we require the Gaussian significance $Z=1.64$. Fig.~\ref{fig:n95} shows $N_{95}$ as a function of $C_2$ and $m$.

\clearpage

\clearpage
\end{document}